\documentclass[epj]{svjour}
\usepackage{graphicx,float,amsmath,amssymb}
\newfont{\ensmathquatorze}{msbm10 scaled 1400}
\newfont{\ensmathonze}{msbm10 scaled 1100}
\newfont{\ensmathdix}{msbm10}
\newfont{\ensmathneuf}{msbm10 scaled 833}
\newfont{\ensmathhuit}{msbm10 scaled 694}
\newfam\ensmathfam                        
\textfont\ensmathfam=\ensmathonze        
\scriptfont\ensmathfam=\ensmathdix       
\scriptscriptfont\ensmathfam=\ensmathhuit

\def\ensmf{\fam\ensmathfam\ensmathonze}         
\def\lsim{\hbox{\lower .8ex\hbox{$\, \buildrel < \over \sim\,$}}}
\def\gsim{\hbox{\lower .8ex\hbox{$\, \buildrel > \over \sim\,$}}}

\newcommand{\APPROX}[1]{                
   {{\raisebox{-.3cm}{$\textstyle\simeq$}} \atop {\scriptstyle{#1}}}}

\def\ZZ{{\ensmf Z}}                 
\def\RR{{\ensmf R}}                 

\begin{document}
\title{Anderson localization of elementary excitations in a one
	dimensional Bose-Einstein condensate}

\author{N. Bilas and N. Pavloff}
\institute{Laboratoire de Physique Th\'eorique
et Mod\`eles Statistiques,
Universit\'e Paris Sud, b\^at. 100, F-91405 Orsay Cedex, France}

\date{Received: date / Revised version: date}
%
\abstract{ We study the elementary excitations of a transversely confined
Bose-Einstein condensate in presence of a weak axial random potential. We
determine the localization length (i) in the hydrodynamical low energy regime,
for a domain of linear densities ranging from the Tonks-Girardeau to the
transverse Thomas-Fermi regime, in the case of a white noise potential and
(ii) for all the range of energies, in the ``one-dimensional mean field
regime'', in the case where the randomness is induced by a series of randomly
placed point-like impurities. We discuss our results in view of recent
experiments in elongated BEC systems.
\PACS{
{03.75.Kk}{Dynamic properties of condensates; collective and
    hydrodynamic excitations, superfluid flow} \and 
{05.60.Gg}{Quantum
    transport} 
} 
} 
\authorrunning{N. Bilas and N. Pavloff}
\titlerunning{Localization of excitations in a one dimensional
Bose-Einstein condensate}

\maketitle


\section{Introduction}

The rapid developments of coherent atom manipulation which has recently
allowed to study atomic interferometry of Bose-Einstein condensate (BEC) on a
chip \cite{Wan05,Shi05} opens up the prospect of considering a whole set of
new transport phenomena in BEC systems. This can be considered as a new domain
for studying the concepts issued form mesoscopic physics. As for the clean 2D
electronic devices considered in this latter field, the BECs are genuinely
phase coherent. Moreover, whereas interactions are difficult to model in
mesoscopic physics, their effects in BEC systems are rather well understood
and are expected to lead to a whole body of interesting phenomena: atom
blockade \cite{Car99}, perfect solitonic-like transmission over a barrier
\cite{Leb03}, non linear resonant transport \cite{Pau05}, breakdown and
revival of Bloch oscillations \cite{Wit05}, to mention just a few examples.

Coherent transport phenomena are of special interest in presence of
disorder. Interference effects have then a prominent role, resulting, in the
non interacting case, in weak or strong localization, as observed in many
different fields (electronic or atomic physics, acoustics or
electromagnetism). The influence of interaction on this phenomenon are of
great interest (see, e.g., the review \cite{Gre92}) and have recently been
addressed in the case of repulsive two body effective interaction for BEC
systems in Refs. \cite{Bil05,Pau05b}. In these latter two references,
interaction effects have been shown to lead to genuinely non-linear phenomena
that profoundly alter the usual picture of Anderson localization.

In the present work, we also consider the influence of interaction on Anderson
localization, but remaining at a linear level, by studying the propagation of
elementary excitations in a disordered BEC system. These are small
deformations of a static background and they can be --at leading order--
described in a linear framework (neglecting phenomena such as Beliaev damping).
Interaction has nonetheless 
a prominent effect on the spectrum of elementary excitations,
which is phonon-like at small energy and becomes similar to the one of non
interacting 
particles at high energy. The crossover between these two regimes occurs at an
energy $\hbar\omega$ of order of the chemical potential $\mu$ of the system.

\includegraphics*[width=7.9cm]{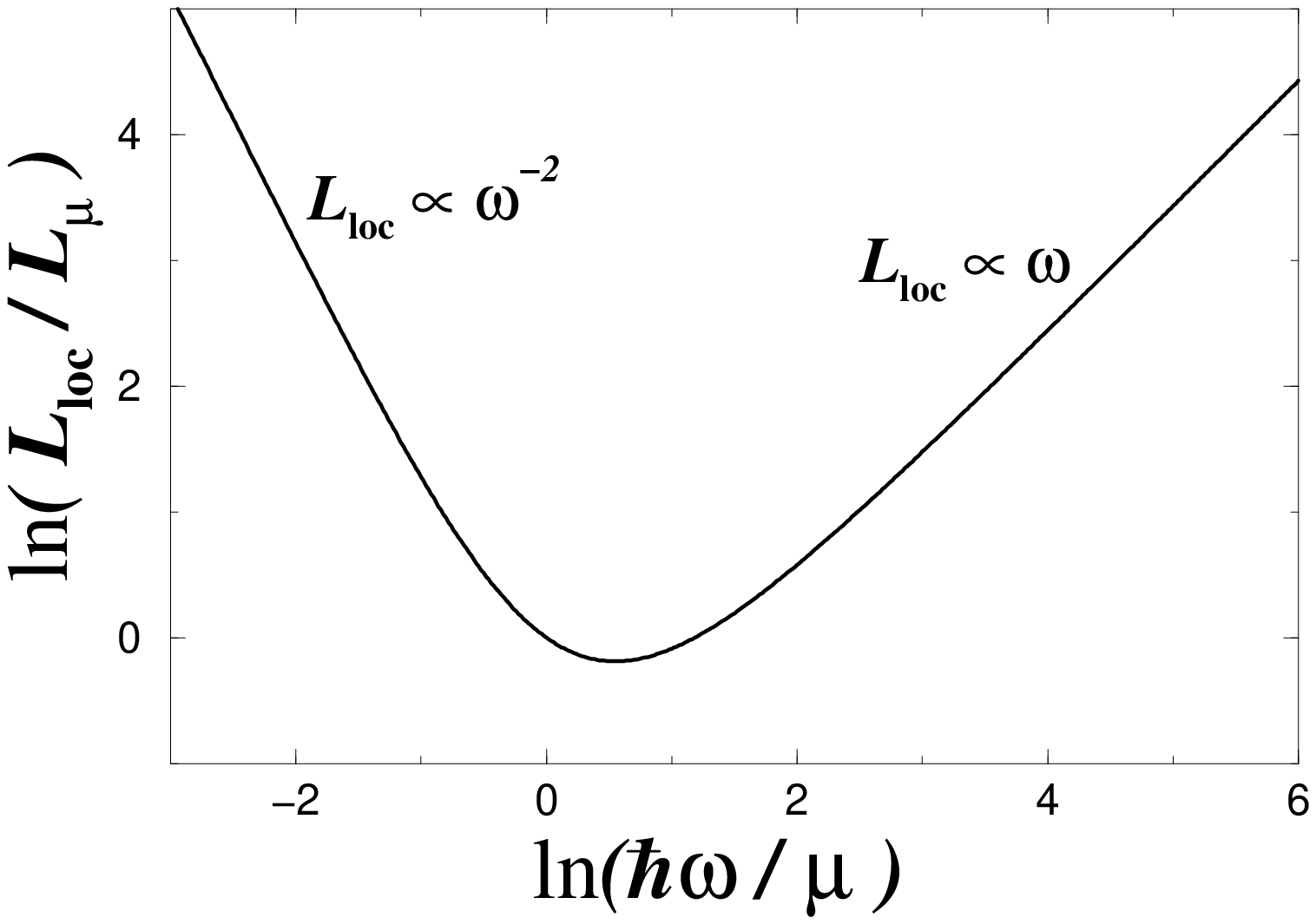}

\vspace*{-0.5 cm}
\noindent Figure 1: {\sl $L_{\rm loc}$ as a function of the energy
	$\hbar\omega$ of an elementary excitation in logarithmic scale ($\mu$ is the
	chemical potential of the system, and $L_\mu$ is the value of $L_{\rm loc}$
	when $\hbar\omega=\mu$). The curve has been drawn within the model used in
	Section \ref{transfer}, employing formulas (\ref{t7b}) and (\ref{b11}). This
	yields $\lambda^2\,n_{\rm imp} L_{\rm
	loc}=4[(\hbar\omega/\mu)^2+1]/[\sqrt{(\hbar\omega/\mu)^2+1}-1]$ (the meaning
	of the parameters $\lambda$ and $n_{\rm imp}$ in this formula is explained
	in Section \ref{transfer}).}

Accordingly, the localization length $L_{\rm loc}$ of the elementary
excitations (i.e., the typical extend of a localized mode, see Section 2 below)
is expected to be similar to the one of phonons at low energy
($\hbar\omega\ll\mu$), and to the one of non interacting particles at high energy
($\hbar\omega\gg\mu)$ \cite{Shl05}. The localization length of non interacting
particles
scales linearly with the energy (at high enough energy, see, e.g.,
\cite{Lif88}), whereas phonons in 1D disordered system have a localization
length which diverges as $\omega^{-2}$ at small $\omega$, as typically observed
in mo\-dels of disordered harmonic chains \cite{Ish73}, in random layered media
\cite{Bal85}, or in continuous models with random elastic properties
\cite{She86}.
Hence, the localization length $L_{\rm loc}$ of the elementary
excitations has the behavior illustrated in Figure 1, with a minimum at
$\hbar\omega\simeq\mu$. The main purpose of the present work is to explicitly
derive this type of behavior within several approximation schemes and
different models of disorder.

The paper is organized as follows. In Section \ref{model} we briefly present
the model and the parameter range in which we are working, together with the
Bogoliubov-de Gennes equations governing the dynamics of the elementary
excitations. In Section \ref{hydro} we consider the large wave-length limit
within an hydrodynamical approach. We consider a Gaussian white noise
potential and show in particular that in this domain, one obtains a
$\omega^{-2}$ behavior of $L_{\rm loc}$. In Section \ref{transfer} we consider
an other type of disorder (randomly placed delta impurities) and work within
the transfer matrix approach. In this regime we are able to work for all the
range of energies and obtain an analytic expression for $L_{\rm loc}$ in the
scarce impurities limit. This expression matches at low energy the one
obtained in Section \ref{hydro} within the hydrodynamical approach. Very
interesting recent experiments have addressed the issue of transport in a
disordered BEC \cite{Lye05,Cle05,Sch05} and in Section \ref{conclusion} we
discuss the relevance of our approach for analyzing some of the experimental
results. Finally, some technical points are given in the Appendices. Appendix
\ref{dos} is devoted to the derivation of a formula allowing to determine the
density of state within the ``phase formalism'' employed in Section
\ref{hydro}. In Appendix \ref{1impurity} we compute the transmission
coefficient of an elementary excitation of energy $\hbar\omega$ over a single
delta-like impurity.

\section{The model}\label{model}
In this Section we present the basic equations describing the elementary
excitations of a one dimensional (1D) Bose-Einstein condensed gas in presence
of disorder. The condensate is formed by atoms of mass $m$ which interact {\it
via} a two-body potential characterized by its 3D s-wave scattering length
$a>0$. The gas is confined to one dimension by a transverse parabolic
potential of frequency $\omega_\perp$ and ``oscillator length''
$a_\perp=(\hbar/m\omega_\perp)^{1/2}$. There is no confinement in the axial
($x$) direction, but disorder is induced along the axis of the guide through a
random potential $U(x)$ whose properties will be specified in the next
sections.

In this Section (and also in Section \ref{transfer}) we restrict ourselves to
the ``1D mean field regime'' \cite{Men02} corresponding to a density range
such that
\begin{equation}\label{e1}
(a/a_\perp)^2 \ll n_{1\rm{D}}\, a \ll 1 \; ,
\end{equation}
\noindent where $n_{1\rm{D}}$ denotes a typical order of magnitude of the 1D
density $n(x,t)$ of the system. The first of the inequalities (\ref{e1})
ensures that the system does not get in the Tonks-Girardeau limit and the
second that the transverse wave function is the ground state of the linear
transverse Hamiltonian, see, e.g., the discussion in Refs.~\cite{Men02,Pet04}.
We address the low density case (Tonks-Girardeau limit) and the high
density case (transverse Thomas-Fermi) in Section \ref{hydro}.

In the 1D mean field regime, the field operator is a function $\hat\Psi(x,t)$
which can be decomposed in the usual Bogoliubov way in c-number (the
superfluid order parameter) plus small terms describing the contribution of
the elementary oscillations (see, e.g., Ref. \cite{Pit03}, chap. 5). For a
stationary condensate, the order parameter is of the form $\psi(x)\exp\{-{\rm
i}\mu t/\hbar\}$ where $\psi(x)$ is real, and the Bogoliubov decomposition
reads
\begin{eqnarray}\label{e2}
\hat\Psi(x,t)& = & {\rm e}^{-{\rm i}\mu t/\hbar}\,\Big\{
\psi(x) +  \nonumber \\
& & 
\sum_\nu[u_\nu(x)\,\hat{b}_\nu\,{\rm e}^{-{\rm i}\omega_\nu t}
+v_\nu^*(x)\,\hat{b}^\dagger_\nu\,{\rm e}^{{\rm i}\omega_\nu t}]
\Big\}\; ,
\end{eqnarray}
\noindent where $\hat{b}_\nu$ and $\hat{b}^\dagger_\nu$ are, respectively, the
annihilation and creation operator of the $\nu$th elementary excitation. In
the following, we drop the subscript $\nu$ for legibility.
The order parameter verifies the Gross-Pitaevskii equation
\begin{equation}\label{e3}
-\frac{\hbar^2}{2m}
 \frac{{\rm d}^2 \psi}{{\rm d} x^2} + \Big\{ U(x) +
g_{1{\rm D}}\,\psi^2(x) \Big\} \psi(x) 
=\mu\,\psi(x) \; ,
\end{equation}
\noindent with $g_{1{\rm D}}=2\hbar \omega_{\perp} a$
\cite{Ols98,Jac98,Leb01}. The functions $u(x)$ and $v(x)$ are solutions of the
Bogoliubov-de Gennes equations (see, e.g., Ref. \cite{Pit03}, chap. 5)
\begin{equation}\label{e4}
\left(\begin{array}{ccc} H &\,& g_{1{\rm D}}\,\psi^2 \\ 
-g_{1{\rm D}}\,\psi^2 &\,& -H\end{array}\right)
\left(\begin{array}{c} u \\ v \end{array}\right) 
= \hbar\omega
\left(\begin{array}{c} u \\ v \end{array}\right) 
\; ,
\end{equation}
where
\begin{equation}\label{e5}
H=-\frac{\hbar^2}{2m}
 \frac{{\rm d}^2}{{\rm d} x^2} + U(x) +2\,g_{1{\rm D}}\,\psi^2(x)-\mu \; .
\end{equation}
In presence of a single elementary excitation of pulsation $\omega$ the
density reads $n(x,t)=|\psi(x)|^2+\delta n(x,t)$ where
the density oscillation is, at leading order:
\begin{equation}\label{e6}
\delta n(x,t)=\psi(x) [u(x) + v(x)]\,{\rm e}^{-{\rm i}\omega t} +
\rm{c.c.} \; ,
\end{equation}
\noindent where ``c.c.'' stands for ``complex conjugate''. In Section
\ref{hydro} we use the notation $\delta n(x)$ for the quantity $\psi(x) [u(x)
	+ v(x)]$. 

In the absence of potential $U$, the order parameter is a constant
$\psi(x)=n_0^{1/2}$ with $\mu=g_{1{\rm D}}\,n_0$, 
the speed of sound in the system is
$c_0=(\mu/m)^{1/2}$ and the healing length is $\xi=\hbar/(m\,c_0)$.

Disorder is induced along the axis $x$ of the guide through the random
potential $U(x)$. Denoting $U_{\rm typ}$ the typical value of $|U(x)|$, we
work in the limit $U_{\rm typ}\ll\mu$. This regime is easily reached
experimentally \cite{Lye05,Cle05} and is very relevant for our purpose because
it corresponds to a range of parameters where Anderson localization is not
blurred by effects connected to ``fragmentation of the condensate''
\cite{corrug}.

In a 1D disordered system the excitations are expected to be localized around
a point with an envelop decreasing exponentially with the distance to this
point. This corresponds to functions $u$, $v$ and $\delta n$ behaving as
$\exp\{\pm\gamma x\}$ when $|x|\to\infty$. $\gamma$ is a function of $\omega$
known as the Lyapunov exponent; it characterizes the localization properties
of the system. Its inverse $L_{\rm loc}=\gamma^{-1}$ is the localization
length \cite{Lif88}. We determine the Lyapunov exponent of the
system in Section \ref{hydro} in the hydrodynamical regime $\hbar\omega\ll \mu$.

In Section \ref{transfer} we approach the problem in a different --but
equivalent-- manner. The disordered potential is assumed to be non zero only
in a finite region of space, between $x=0$ and $L$. We consider an elementary
excitation of pulsation $\omega$ incident on the random
potential. The corresponding transmission coefficient $T$ through the
disordered region is related to the Lyapunov exponent {\it via}
$\gamma=-\frac{1}{2}\lim_{L\to\infty} L^{-1}\ln T$ \cite{Lif88}. This is
simply connected to the fact that the incident wave function decreases
exponentially --at a rate $\gamma$-- in the disordered region, and this
corresponds finally to a transmission probability which is (within logarithmic
accuracy) $T \sim \exp(-2\,\gamma L)$.

We note here important features of the localization properties of the
elementary excitations. First, Eq.~(\ref{e4}) admits a zero energy solution
for $u(x)=-v^*(x)=\psi(x)$. Thus, whatever the disordered potential $U(x)$,
the excitation at $\omega=0$ is delocalized since $\psi(x)$ extends to
infinity. This implies that $L_{\rm loc}$ diverges as $\omega\to 0$. Secondly,
at $\omega\to \infty$ the high energy part of the spectrum is well described
by a single particle description obtained by neglecting the coupling between
the positive ($u$) and negative ($v$) frequency components of the excitations
(see, e.g., Ref. \cite{Pit03}, chap. 12). In this limit one can set $v=0$ in
Eq.~(\ref{e4}) and the system is described by the Schr\"odinger-like
Hamiltonian $H$ (\ref{e5}) which localization length behaves as $L_{\rm loc}
\propto \omega$ at high energy. Thus, as already anticipated in the
introduction, we expect a behavior of $L_{\rm loc}$ similar to what has been
drawn in Fig. 1.

\section{Hydrodynamical approach: $\hbar\omega\ll \mu$}\label{hydro}

The results obtained in this Section are derived within the 1D mean field regime
(\ref{e1}). As explained at the end of the Section, they can be easily
generalized in the transverse Thomas-Fermi regime and even in the
Tonks-Girardeau limit.

In the present Section we only consider the low frequency excitations
($\hbar\omega\ll\mu$). These involve large wave lengths (which are of order
$2\pi\,c_0/\omega$, when $\omega\to 0$) and accordingly, features at small length
scale are not relevant in the potential seen by the excitations. In
particular, the ground state order parameter can be evaluated in the
Thomas-Fermi approximation \cite{TFnote} leading to
\begin{equation}\label{h1}
\psi(x)=\sqrt{\frac{\mu-U(x)}{g_{1{\rm D}}}} \; .
\end{equation}
By reintroducing this ansatz in Eq.~(\ref{e3}), one can easily show (provided
$U_{\rm typ}$ is small compared to $\mu$) that the Thomas-Fermi result
(\ref{h1}) is valid in the limit $\xi\ll r_c$, where $r_c$ fixes the length
scale of typical variations of $U$ (for instance this is the correlation
length of the random potential). If besides, one considers the limit $\xi\ll
c_0/\omega$, the density oscillations $\delta n(x)$ obey the hydrodynamical
equation \cite{Str96,Ohb97}
\begin{equation}\label{h2}
-\omega^2 \delta n(x) = 
\frac{\rm d}{{\rm d}x}\left(
c^2(x) \frac{\rm d}{{\rm d}x}\delta n(x) 
\right) \; ,
\end{equation}
\noindent where $c(x)=\{[\mu-U(x)]/m\}^{1/2}$ is a local sound velocity.

Disorder is induced along the axis of the guide through the random
potential $U(x)$ which is assumed to have zero mean. The case $\langle U
\rangle \ne 0$ can be treated with a trivial extension of the present approach
which is explained at the end of the Section. In the following of this
Section, $U$ will be approximated by a Gaussian white noise. The hypothesis of
white noise is only valid if the wave length of the excitations is large
compared to the correlation length $r_c$ of the true $U$ (which is not a
perfect white noise if we want the Thomas Fermi approximation (\ref{h1}) to
hold). Hence, in the present Section, we make the consistent hypothesis that
\begin{equation}\label{h3}
\xi \ll r_c \ll \frac{2\,\pi\,c_0}{\omega} \; .
\end{equation}
When the inequality (\ref{h3}) is verified, Equations (\ref{h1}) and
(\ref{h2}) are both valid and furthermore the approximation of the random
potential by a white noise is sound. In the following we thus write
\begin{equation}\label{h4}
\langle U(x)U(0)\rangle= 
\left(\frac{\hbar^2}{m}\right)^2 \! D\,\delta(x) \; .
\end{equation}
We now evaluate the localization length corresponding to Eq.~(\ref{h2}) by
means of the phase formalism (see Ref.~\cite{Lif88}). We consider a real
solution of (\ref{h2}) and define the functions $\alpha(x)$ and $\beta(x)$ by
\begin{equation}\label{h5}
\alpha(x)=\frac{\delta n(x)}{\delta n^*} \; ,\quad
\beta(x)=-\frac{c^2(x)}{c_0\,\omega}\,
\frac{{\rm d}\alpha}{{\rm d}x} \; .
\end{equation}
In (\ref{h5}) the quantity $\delta n^*$ is a typical value of $\delta n(x)$
which is introduced for dimensional purpose, but plays no role in the
following [since Eq.~(\ref{h2}) is linear]. The functions $\alpha$ and $\beta$
satisfy the following system of equations:
\begin{equation}\label{h5b}
\frac{{\rm d}\alpha}{{\rm d}x}=-\frac{\omega}{c_0}
\Big[1+\eta(x)\Big]\beta(x)\; ,\quad
\frac{{\rm d}\beta}{{\rm d}x}=\frac{\omega}{c_0}\,\alpha(x) \; .
\end{equation}
In the first of equations (\ref{h5b}), the term $\eta(x)$ is equal to
$U(x)/[\mu-U(x)]$. In all the following we assume that $U_{\rm typ}$ 
is much smaller than $\mu$, and we write $\eta(x)\simeq U(x)/\mu$
\cite{rigor}.

It is convenient to parametrize the functions $\alpha$ and $\beta$ 
in the form
\begin{equation}\label{h6}
\alpha(x)=r(x)\cos\theta(x) \; ,\quad
\beta(x)=r(x)\sin\theta(x)\; .
\end{equation}
The functions $\theta(x)$ and  $r(x)$
describe respectively the phase and the envelope of the density
oscillations $\delta n(x)$ [and accordingly of $u(x)$ and of $v(x)$]. In
particular, the Lyapunov exponent is defined by 
\begin{equation}\label{h7}
\gamma(\omega)=\lim_{x\to\infty} \frac{\langle\ln r(x)\rangle}{x} \; .
\end{equation}
It is convenient to introduce the quantity $z=\alpha/\beta$ because, owing to the
equality
\begin{equation}\label{h11}
\ln r^2(x)=\frac{2\omega}{c_0}\int_0^x\!\!\!\! z(x')\, {\rm d}x'+
\ln\beta(0)
- \ln\sin^2\theta(x) \; ,
\end{equation}
\noindent and to the fact that the probability density of $\sin\theta$ (and
thus also that of $z=\cot\theta$) becomes stationary (i.e., $x$ independent)
at large $x$ \cite{Lif88}, one can write
\begin{equation}\label{h12}
\gamma=\frac{\omega}{c_0}\lim_{x\to\infty} 
x^{-1}\int_0^x\!\! \langle z(x')\rangle \, {\rm d}x'
=\frac{\omega}{c_0}\, \langle z\rangle_{\rm st} \; ,
\end{equation}
\noindent where $\langle z\rangle_{\rm st}$ is the mean value of $z$ in the
stationary regime. This quantity is determined as follows.
From (\ref{h5b}) one sees that $z$ verifies the following
stochastic differential equation
\begin{equation}\label{h10}
-\frac{c_0}{\omega}\,\frac{{\rm d}z}{{\rm d}x}=1+z^2+\frac{U(x)}{\mu} \; .
\end{equation}
Let $P(z;x)dz$ be the probability that $z(x)$ lies in the interval $z$,
$z+dz$. From (\ref{h10}) and (\ref{h4}) $P$ verifies the
Fokker-Planck equation (see, e.g., \cite{Lif88,Itz89})
\begin{equation}\label{h13}
\frac{\partial P}{\partial x}=\frac{\omega}{c_0}\,
\frac{\partial}{\partial z}\left[
(1+z^2)\,P+\frac{\omega\delta}{2}\frac{\partial P}{\partial z}\right] \; ,
\end{equation}
\noindent where $\delta=\xi^4\, D/c_0$. The stationary regime corresponds to
the case where $\partial_xP=0$. In this case, writing $P=P_{\rm st}(z)$,
Eq. (\ref{h13}) yields
\begin{equation}\label{h13b}
(1+z^2)\,P_{\rm st}+
\frac{\omega\,\delta}{2}\frac{{\rm d} P_{\rm st}}{{\rm d} z}=J_\omega\; ,
\end{equation}
where $J_\omega$ is an integration constant. The solution of (\ref{h13b}) is
\begin{eqnarray}\label{h14}
& P_{\rm st}(z)& = \frac{2\,J_\omega}{\omega\,\delta} 
\int_0^{+\infty}\!\!\!\!\!\! {\rm d}t \nonumber \\
& & 
\exp\left\{
\frac{2}{\omega\delta}\left[-(1+z^2)t+z t^2-\frac{t^3}{3}\right]
\right\}\; .
\end{eqnarray}
The value of $J_\omega$ is
fixed by the normalization of $P_{\rm st}$. One obtains
\begin{equation}\label{h15}
J_\omega^{-1}=\sqrt{\frac{2 \pi}{\omega \delta}}
\int_\RR\!\!\exp\left[-\frac{12\,t^2+t^6}{6\, \omega \delta} \right]
\,{\rm d}t \; .
\end{equation}
Simple algebra allows to express the average $\langle z\rangle_{\rm st}=$ \\
$\int_\RR z\, P_{\rm st}(z)\, {\rm d}z$ under the following form:
\begin{equation}\label{h16}
\langle z\rangle_{\rm st}=J_\omega\sqrt{\frac{\pi}{2\omega\delta}}\,
\int_\RR\!\!
\exp\left[-\frac{12\,t^2+t^6}{6\, \omega \delta}\right]\,t^2\,{\rm d}t 
\; .
\end{equation}
We are primarily interested in this Section 
in the small frequency evaluation of the
Lyapunov exponent, because Eq. (\ref{h2}) is expected to describe the elementary
excitations only in the domain $\hbar\omega/\mu\ll 1$.
An expansion of the integrals
(\ref{h15}) and (\ref{h16}) in the limit $\omega\delta\to 0$ yields, after
reinserting in (\ref{h12}):
\begin{equation}\label{h17}
\gamma=\frac{\xi^2\,D}{8}\left(\frac{\hbar\omega}{\mu}\right)^2\,
\left[1-\frac{15}{16}\left(\frac{\omega\delta}{2}\right)^2+\cdots \right]\; .
\end{equation} 
Although the high frequency limit is not expected to be relevant in the
hydrodynamical regime, we note for completeness that when
$\omega\delta\to\infty$ one obtains
\begin{eqnarray}\label{h18}
\gamma& = & \frac{\omega}{c_0}\,\sqrt{\frac{3}{8\pi}}\,
\Gamma\Big(\frac{5}{6}\Big)
\left(\frac{\omega\delta}{\sqrt{6}}\right)^{1/3}\times\nonumber \\
& & \left[1-
\frac{\Gamma\!\left(\frac{5}{6}\right)}{\sqrt{\pi}}
\left(\frac{\sqrt{6}}{\omega\delta}\right)^{2/3} + \right. \nonumber \\
& & \left. \left(
\frac{2\sqrt{\pi}}{3\Gamma\!\left(\frac{5}{6}\right)}
-\frac{2[\Gamma\!\left(\frac{5}{6}\right)]^2}{\pi}
\right)
\left(\frac{\sqrt{6}}{\omega\delta}\right)^{4/3}
+ \cdots
\right]\; .
\end{eqnarray} 

The exact value of $\gamma$ --as determined numerically from Eqs. (\ref{h12}),
(\ref{h15}) and (\ref{h16})-- is represented in Fig. 2 (solid line).

\includegraphics*[width=8cm]{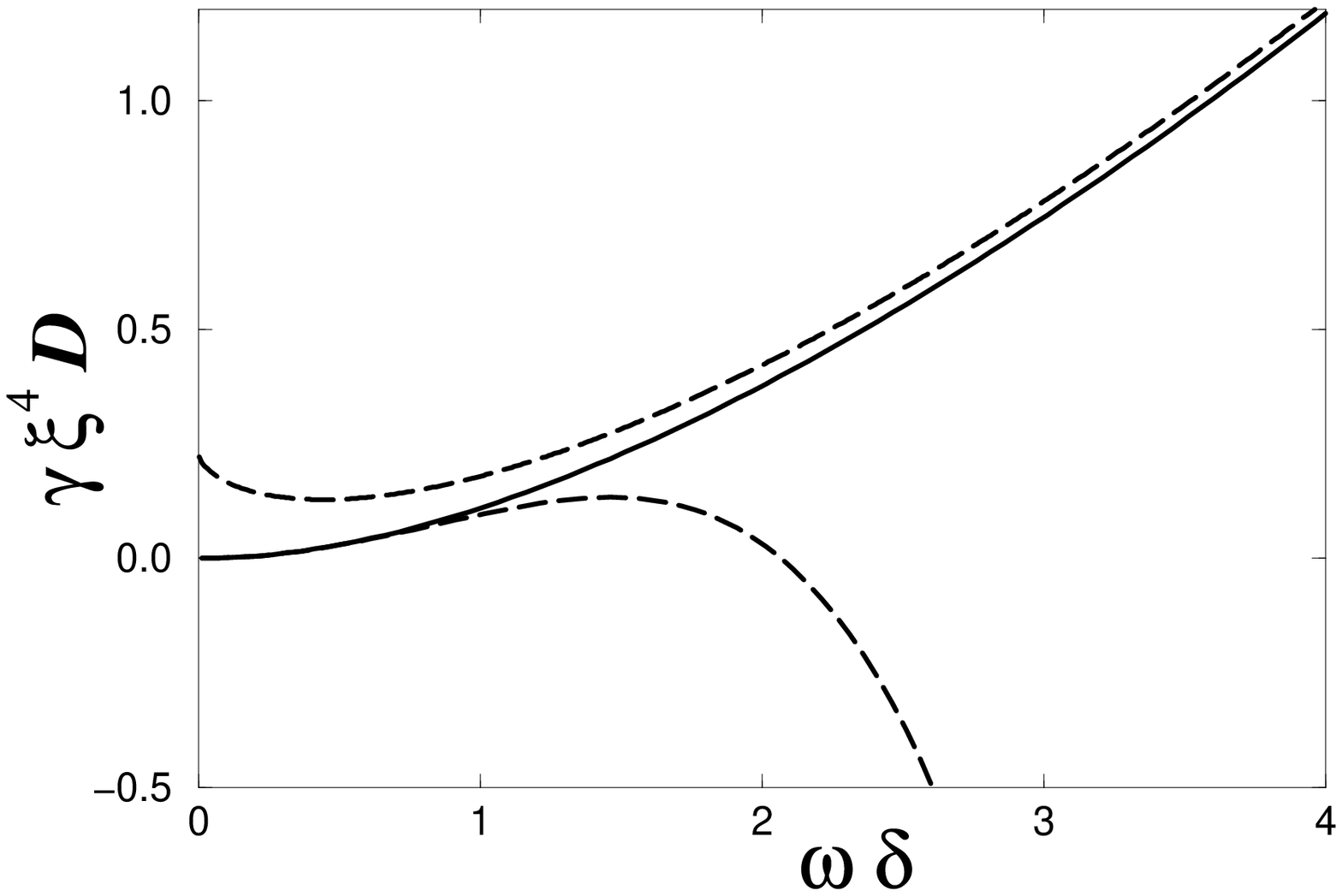}

Figure 2: {\sl $\gamma$ as a function of $\omega$ in rescaled units. The solid
	line is the numerical evaluation of $\gamma$ using formulas (\ref{h12}) and
	(\ref{h16}). The dashed lines are the small and large $\omega\delta$
	approximations [Eqs. (\ref{h17}) and (\ref{h18})].}

\

The quantity $J_\omega$ is also of interest for itself, because it gives
informations on the density of states of the excitations. It is show in
Appendix \ref{dos} that, if $N(\omega)$
denotes the integrated density of state per unit
length, one has
\begin{equation}\label{h19}
N(\omega)=\frac{\omega}{c_0}\, J_\omega \; .
\end{equation}
From (\ref{h15}) one gets the following expansions:
\begin{eqnarray}\label{h21} 
N(\omega) & = &
\frac{3\,\omega}{c_0}\,
\frac{\Gamma(\frac{5}{6})}{(2\pi)^{3/2}}
\left(\frac{\omega\delta}{\sqrt{6}}\right)^{1/3}\times\nonumber \\
& & \left[1+
\frac{\Gamma\!\left(\frac{5}{6}\right)}{\sqrt{\pi}}
\left(\frac{\sqrt{6}}{\omega\delta}\right)^{2/3}+\cdots
\right]\; .
\end{eqnarray}
\noindent when $\omega\delta\gg 1$, and
\begin{equation}\label{h20} 
N(\omega)=\frac{\omega}{\pi\,c_0} \, \left[ 1 + \frac{5}{32}
	\left(\frac{\omega\delta}{2}\right)^2 + \cdots \right] \; ,
\end{equation}
\noindent when $\omega\delta\ll 1$. In the relevant regime of low excitation
energies, the leading order in (\ref{h20}) coincides the result in absence of
disorder, where one has a linear dispersion relation $\omega=c_0|q|$ in the
hydrodynamical regime. This confirms what could have been already anticipated
from the fact that $\gamma\to 0$ when $\omega\to 0$: the low lying excitations
are poorly affected by the presence of disorder (the relevant small parameter
being $\omega\delta$). In particular, there is no trapping of the elementary
excitations by the disorder and no Lifshitz tail in the density of state. This
is linked to the fact that $\omega=0$ constitutes what is called a ``stable
genuine boundary of the spectrum'' in the book by Lifshits, Gredeskul and
Pastur (see Ref.~\cite{Lif88}, section 7.3).

The results presented in this Section have been obtained for a random potential
with zero mean. They are very easily adapted to the case $\langle U\rangle\ne
0$: it suffices to write $U(x)=\langle U\rangle+U_1(x)$, and to define
$\mu_1=\mu-\langle U\rangle$, $c_1=(\mu_1/m)^{1/2}$, $\xi_1=\hbar/mc_1$,
$\delta_1=\xi^4_1 D/c_1$. Then, all the results presented from Eq. (\ref{h5}) to
Eq. (\ref{h20}) remain valid provided $U(x)$, $\mu$, $c_0$, $\xi$ and
$\delta$ are replaced by the similar quantities with subscript ``1'', with the
coefficient $D$ being now defined by $\langle
U_1(x)U_1(0)\rangle=(\hbar^2/m)^2\,D\,\delta(x)$ [instead of (\ref{h4})].

  The present hydrodynamical approach is very interesting because it has
natural extensions out of the 1D mean field regime defined by Eq. (\ref{e1}).
For high linear densities, when $n_{1\rm{D}}a\gg 1$, one reaches the
``transverse Thomas-Fermi regime'' also named ``3D cigar'' in Ref.
\cite{Men02}. In this regime the system cannot be considered as truly
uni-dimensional. However, the lowest branch of the spectrum corresponds to
excitations that are isotropic in the transverse direction, and, as shown by
Stringari in Ref. \cite{Str98}, they can still be described within the
hydrodynamical approach. In this case, averaging the 3D hydrodynamical
equation over the transverse direction, one gets a 1D equation
of the form (\ref{h2}) where the local sound velocity $c(x)$ is now taken
to be $c(x)=\{[\frac{1}{2}\mu-U(x)]/m\}^{1/2}$. So, all the results
presented from Eq. (\ref{h5}) to Eq. (\ref{h20}) remain valid provided $\mu$,
$c_0$ and $\xi$ and are replaced by $\mu'=\mu/2$, $c_0'=(\mu'/m)^{1/2}$ and
$\xi'=\hbar/(m\,c_0')$.

  The low density regime $n_{1\rm{D}}a \ll (a/a_\perp)^2$ (Tonks-Gi\-rar\-deau)
can also be studied within the hydrodynamical framework (see for instance Ref.
\cite{Men02}). In this case one has $\mu=(\pi\,\hbar\, n_0)^2/2m$,
$c_0=(2\mu/m)^{1/2}$ and Eq.~(\ref{h2}) is replaced by
\begin{equation}\label{h22}
-\omega^2 \delta n(x) = 
\frac{\rm d}{{\rm d}x}\left\{
c(x)\frac{\rm d}{{\rm d}x} \Big[ c(x)\,\delta n(x)\Big] 
\right\} \; ,
\end{equation}
with the local sound velocity being defined by
$c(x)=\{[\frac{2}{m}[\mu-U(x)]\}^{1/2}$. In the present case we define
	[instead of (\ref{h5})]
\begin{equation}\label{h23}
\alpha(x)=\frac{c(x)}{c_0}\,\frac{\delta n(x)}{\delta n^*}
\; ,\quad
\beta(x)=\frac{c(x)}{\omega}\,\frac{{\rm d}\alpha}{{\rm d}x} \; .
\end{equation}
Writing $\alpha=r\,\sin\theta$ and $\beta=r\,\cos\theta$, one
obtains:
\begin{equation}\label{h24}
\frac{{\rm d}r}{{\rm d}x}=0 \; ,\quad
\frac{{\rm d}\theta}{{\rm d}x}=\frac{\omega}{c(x)}\; .
\end{equation}
From the second of these equations, in the limit where
$1/c(x)=c_0^{-1}[1+\frac{1}{2}U(x)/\mu]$, one can show that the phase
$\theta(x)$ has a Gaussian distribution of the form
\begin{equation}\label{h25}
Q(\theta;x)=\frac{1}{\sqrt{2\pi\omega^2x \delta/c_0}} \; 
\exp\Big\{-
\frac{(\theta-\theta_0-\frac{\omega x}{c_0})^2}{2\omega^2 x \delta/c_0}
\Big\} \; ,
\end{equation}
\noindent with $\delta=(\hbar^2/2m\mu)^2D/c_0$. 

The first of Eqs. (\ref{h24}) is more interesting. It shows that the envelope of
function $\alpha(x)$ remains exactly constant. Assuming that the localization
properties of $\alpha(x)$ are the same than those of $\delta n(x)$ \cite{beware},
this equation points to the absence of exponential localization in the
hydrodynamical limit of the Tonks-Girardeau regime.

\section{Transfer Matrix approach}\label{transfer}

In this Section, we study Anderson localization of the elementary excitations
of a Bose-Einstein condensate with an other type of disordered potential
and in a framework different from the one used in
the previous Section. Namely, we study the transmission through a disordered
region of extend $L$, in the 1D mean field regime (\ref{e1}),
by means of a transfer matrix approach for a disordered
potential:
\begin{equation}\label{t1}
U(x)=g_{\rm imp}\sum_{n}\delta(x-x_n) \; ,{\rm where}\quad
g_{\rm imp}=\lambda\,\mu\,\xi\; .
\end{equation}
 
$U(x)$ describes a series of static impurities with equal intensity and random
positions $x_n$. The peak intensity is measured by the dimensionless parameter
$\lambda$. We consider here the repulsive case $\lambda>0$.
The $x_n$'s are uncorrelated and uniformly distributed with
mean density $n_{\rm imp}$. In this case $<\!\!U(x)\!\!> =
g_{\mbox{\tiny{imp}}} n_{\rm imp}$ and $<\!\!U(x_1)U(x_2)\!\!> -
<\!\!U(x_1)\!\!>\times<\!\!U(x_2)\!\!> = (\hbar^2/m)^2 D \,\delta(x_1-x_2)$,
with $D=n_{\rm imp}(\lambda/\xi)^2$. From what is known in the case of
Schr\"odinger equation, this type of potential is typical insofar as
localization properties are concerned \cite{Lif88}. Besides, it has recently
been proposed to implement a very similar type of random potential by using
two different atomic species in an optical lattice \cite{Gav05}.

The static background is deformed around each impurity over a distance which
is at most of order $\xi$. We consider the regime where this deformation does
not extend to the nearest impurity ($n_{\rm imp}\xi\ll 1$ \cite{avdist}). In
this case, the propagation of an elementary excitation in presence of the
disordered potential $U(x)$ can be treated as a sequence of scatterings over
isolated perturbations. Besides --as shown in Appendix \ref{1impurity}-- both
the scattering of an elementary excitation over such a perturbation, and
its propagation between two successive impurities (separated by a distance
$\ell$) are, in this regime, described by a $2\times 2$ transfer matrix, denoted
respectively ${\cal T}_\lambda$ and ${\cal T}_0(\ell)$ with (see, e.g.,
\cite{Pen94})
\begin{equation}\label{t2}
{\cal T}_\lambda=\left(
\begin{array}{cc}
1/t^*_\lambda &  -r^*_\lambda/t^*_\lambda\\
-r_\lambda/t_\lambda  &  1/t_\lambda\end{array}
\right) \; , \;
{\cal T}_0(\ell)=\left(
\begin{array}{cc}
1/t_0^* &  0\\
0  &  1/t_0\end{array}
\right) \; .
\end{equation}
$r_\lambda$ and $t_\lambda$ in Eq. (\ref{t2}) are the transmission and
reflexion amplitudes of an elementary excitation with energy $\hbar\omega$
across the background deformation induced by a single delta-like impurity.
Their dependence on $\lambda$ and $\omega$ is determined in Appendix
\ref{1impurity} [Eqs. (\ref{b7}) and (\ref{b8})]. The scattering states we
choose for writing the matrices ${\cal T}_\lambda$ and ${\cal T}_0(\ell)$ are
the one introduced in this Appendix. They are pictured in Fig. 3.

\

\begin{center}
\begin{picture}(8,4)
\put(2.5,0){\frame{\includegraphics*[width=2.8cm]{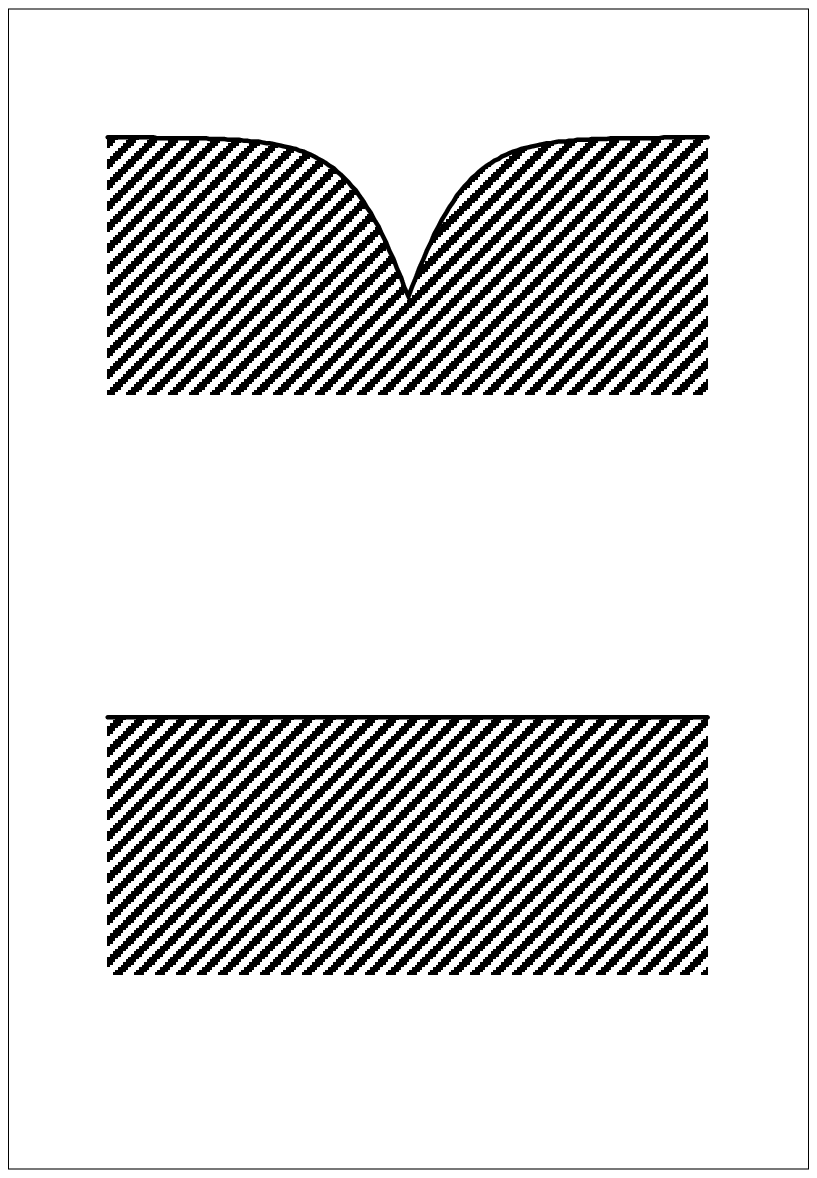}}}
\put(3.8,2){or}

\put(0,3.5){$e^{i q (x-x_{\rm in})}
\left(\!\!\begin{array}{c} u_\omega^* \\ v_\omega^*\end{array}\!\!\right)$}
\put(0,3){\vector(1,0){1.5}}

\put(0,1){$e^{i q (x_{\rm in}-x)}
\left(\!\!\begin{array}{c} u_\omega \\ v_\omega\end{array}\!\!\right)$}
\put(1.5,0.5){\vector(-1,0){1.5}}

\put(5.5,3.5){$e^{i q (x-x_{\rm out})}
\left(\!\!\begin{array}{c} u_\omega \\ v_\omega\end{array}\!\!\right)$}
\put(6.0,3){\vector(1,0){1.5}}

\put(5.5,1){$e^{i q (x_{\rm out}-x)}
\left(\!\!\begin{array}{c} u_\omega^* \\ v_\omega^*\end{array}\!\!\right)$}
\put(7.5,0.5){\vector(-1,0){1.5}}

\put(1,2){\line(1,0){1.5}}
\put(2.5,2){\circle*{0.15}}
\put(5.3,2){\circle*{0.15}}
\put(5.3,2){\vector(1,0){1.5}}
\put(2.0,2.2){$x_{\rm in}$}
\put(5.4,2.2){$x_{\rm out}$}
\put(6.8,1.8){$x$}
\end{picture}
\end{center}

Figure 3: {\sl Scattering channels used for writing the transfer matrices
	${\cal T}_\lambda$ and ${\cal T}_0(\ell)$ of (\ref{t2}). In the case of
	unperturbed motion over a length $\ell$ one has $x_{\rm out}-x_{\rm
	in}=\ell$. In the case of scattering by a delta peak located at $(x_{\rm
	in}+x_{\rm out}) / 2$, one should take $\xi \ll x_{\rm out}-x_{\rm in} \ll
	n_{\rm imp}^{-1}$.}

\

The coefficients $u_\omega$ and $v_\omega$ in Figure 3 are chosen in order to
make the incoming and outgoing channels identical to these appearing
naturally in Appendix \ref{1impurity} when considering the scattering of an
elementary excitation by a single impurity. One thus takes
\begin{equation}\label{t3}
\left(
\begin{array}{c}
u_\omega\\
v_\omega
\end{array}\right)
=
\left(
\begin{array}{c}
\left[\frac{q \xi}{2}+\frac{\omega}{c_0 q}+i\right]^2 \\
\left[\frac{q \xi}{2}-\frac{\omega}{c_0 q}+i\right]^2
\end{array}\right) \; .
\end{equation}
\noindent where $q$ is defined in Eq. (\ref{b5}). In the case of scattering by
an impurity, this corresponds indeed to the scattering channels defined by Eqs.
(\ref{b3}), (\ref{b4}) and (\ref{b6}). In the case of free motion over a
length $\ell$, it is easy to see that these scattering channels correspond to
a matrix ${\cal T}_0(\ell)$ such as defined in Eq.~(\ref{t2}) with
\begin{equation}\label{t4}
t_0(\ell,\omega)=e^{i(q\,\ell - 2\,\alpha)} \; , \;\mbox{where}\quad
e^{- 2\,i\,\alpha}=\frac{u_\omega^*}{u_\omega}=\frac{v_\omega^*}{v_\omega} \; .
\end{equation}

Then, the scattering by a series of $N$ delta peaks separated by distances
$\ell_1=x_2-x_1$, ..., $\ell_{N-1}=x_N-x_{N-1}$, is described by the transfer
matrix ${\cal T}_N$ which is the product
\begin{equation}\label{t5a}
{\cal T}_N={\cal T}_\lambda \times {\cal T}_0(\ell_{N-1})\times
{\cal T}_\lambda ... \times {\cal T}_0(\ell_1)
\times {\cal T}_\lambda \; .
\end{equation}
 ${\cal T}_N$ defined in Eq. (\ref{t5a}) is of the general form
\begin{equation}\label{t5}
{\cal T}_N=\left(
\begin{array}{cc}
1/t_N^* &  -r_N^*/t_N^*\\
-r_N/t_N  &  1/t_N\end{array}
\right) \; .
\end{equation}
Eq. (\ref{t5}) is used for computing the reflexion and transmission amplitudes
($r_N$ and $t_N$) of the elementary excitation over the potential (\ref{t1}).
The transmission probability over this potential is $T_N=|t_N|^{2}$.

As discussed at the end of Section 2, the analogous of the Lyapunov exponent
already computed in Section \ref{hydro} [Eq. (\ref{h7})] is here defined as
\begin{equation}\label{t6}
\gamma=-\lim_{N\to\infty}\, 
\frac{n_{\rm imp}}{N} \; \langle \ln|t_N| \rangle
=-\lim_{N\to\infty}\, 
\frac{n_{\rm imp}}{2\,N} \; \langle \ln T_N \rangle
 \; . 
\end{equation}
We calculated $\gamma$ numerically, by a Monte Carlo averaging over 50
realizations of the disorder, taking $N=2000$ \cite{self}. The result is shown
in Figure 4 for $\lambda=1$ and $n_{\rm imp}\,\xi=0.02$. In the present model
the lengths $\ell_i=x_{i+1}-x_i$ are independent, Poisson distributed, random
variables with $P(\ell) = n_{\rm imp}\,\exp\{-\ell\, n_{\rm imp}\}$. Thus, for
a fraction of lengths equal to $n_{\rm imp}\,\xi$ the transfer matrix approach
fails because the distance between two successive impurities is smaller than
$\xi$ \cite{fraction}. This is the reason why we consider a rather
small value of density of impurities: for the chosen value $n_{\rm
imp}\xi=0.02$, only 2 \% of the distances violate the criterion of
applicability of the transfer matrix approach.

\

\includegraphics*[width=8cm]{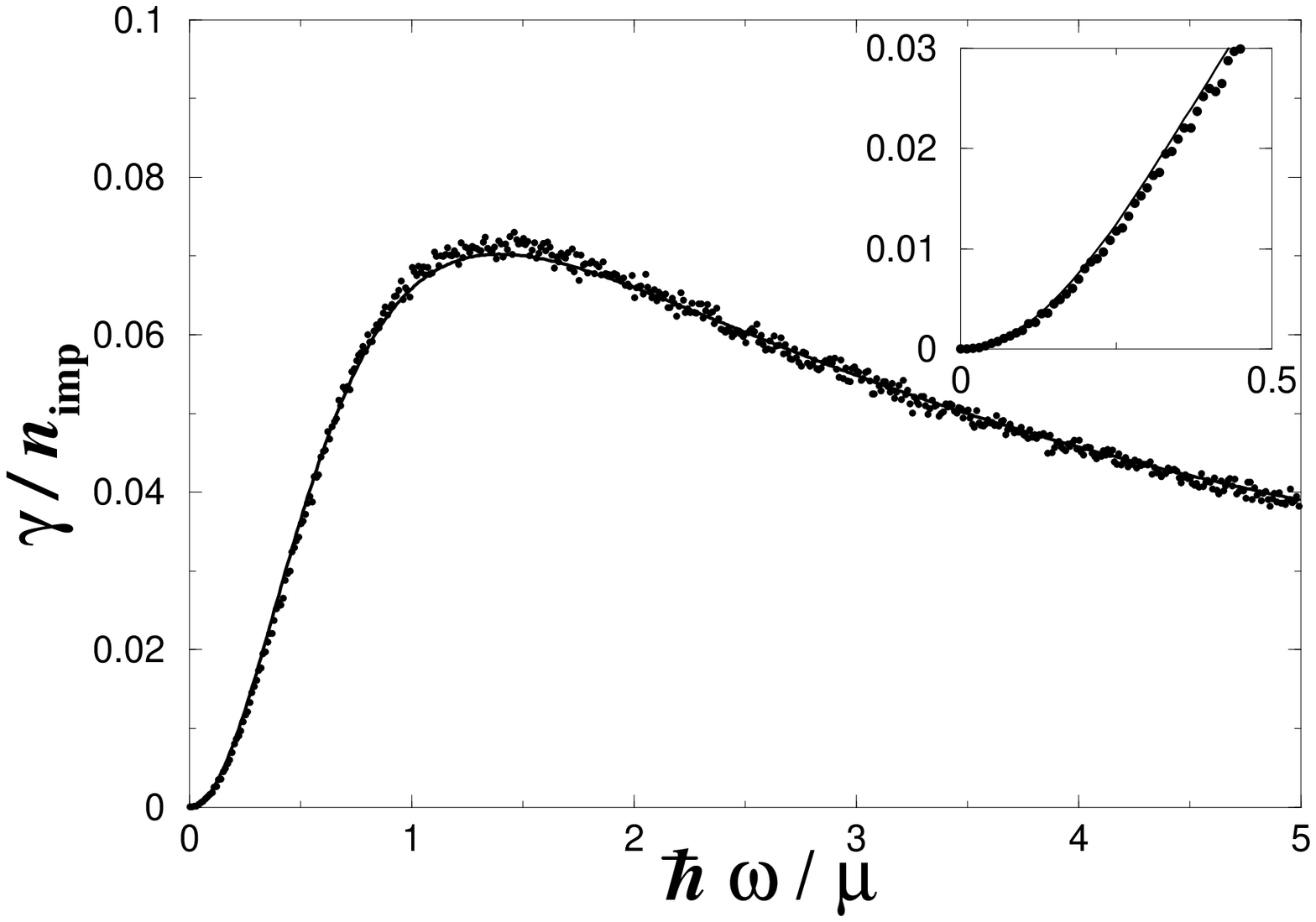}

Figure 4: {\sl $\gamma$ as a function of $\omega$ in rescaled units. The plot
	is drawn for $\lambda=1$ and $\xi\,n_{\rm imp}=0.02$. The dots are the
	results of the numerical simulation and the solid line is the analytical
	result from Eq.~(\ref{t7b}). The inset displays a blowup of the figure at low
	energy.}

\

As shown in Ref. \cite{Ber97}, in the limit $n_{\rm imp}\ll q$, one can
obtain an analytical estimate of $\gamma$. From the relation ${\cal
	T}_{N+1}={\cal T}_\lambda\times {\cal T}_0(\ell_{N})\times {\cal T}_N$ one
gets
\begin{eqnarray}\label{t7}
\langle \ln|t_{N+1}| \rangle & = & \ln|t_\lambda| 
+ \langle \ln|t_N|\rangle\nonumber \\
& - &
\left\langle\ln|1+r_\lambda\,r_N^*\frac{t_N}{t_N^*}\,t_0^2(\ell_N)|\right\rangle
\; .
\end{eqnarray}
$\ell_N$ is typically of order $n_{\rm imp}^{-1}$, and in the limit $n_{\rm
imp}\ll q$, one may assume that the phase of $t_0(\ell_N,\omega)$ given in
(\ref{t4}) is uniformly distributed in $[0,2\pi]$. Then, the last term of the
r.h.s. averages out to zero \cite{And80,Ber97}. This yields
\begin{equation}\label{t7b}
\gamma=- n_{\rm imp}\,\ln|t_\lambda| 
=- \frac{n_{\rm imp}}{2}\,\ln T_\lambda
\; ,
\end{equation}
where we recall that the explicit expression of $t_\lambda$ is
given in Eq. (\ref{b7}) and $T_\lambda=|t_\lambda|^{2}$. 
Formula (\ref{t7b}) corresponds to the solid line in
Fig.~4. The agreement with the result of the numerical simulation is very
good, even at low energy, as shown in the inset of the figure. This is not a
surprise because the breakdown of (\ref{t7b}) is expected only at extremely
low energies for the present value $n_{\rm imp}$: when $q\lsim n_{\rm imp}$,
i.e., $\hbar\omega/\mu\simeq \xi\,q\lsim 0.02$. For larger values of $n_{\rm
imp}$, the good agreement of Eq. (\ref{t7b}) with the numerical data is
limited to a smaller range of energies, mainly because the transfer matrix
approach fails.

From Eq.~(\ref{b10}), in the limit of small $\omega$ and
$\lambda\ll 1$ formula (\ref{t7b}) yields
\begin{equation}\label{t8}
\gamma\simeq 
\frac{\lambda^2}{8}\,\left(\frac{\hbar\,\omega}{\mu}\right)^2 n_{\rm imp} \; .
\end{equation}
The precise range of validity of formula (\ref{t8}) in the energy domain is
expected to be 
$\xi\,n_{\rm imp}\ll \hbar\,\omega/\mu\ll 1$; the first inequality ensures
that (\ref{t7b}) is valid and the second that (\ref{b10}) is applicable.
The accuracy of formula (\ref{t8}) is tested in Fig. 5 in the case
$\lambda=0.2$ and $\xi\,n_{\rm imp}=0.03$. As already seen on Fig. 4, one
notices on this Figure that the restriction $\xi\,n_{\rm imp}\ll
\hbar\,\omega/\mu$ turns out to be of no practical importance.

\

\includegraphics*[width=8cm]{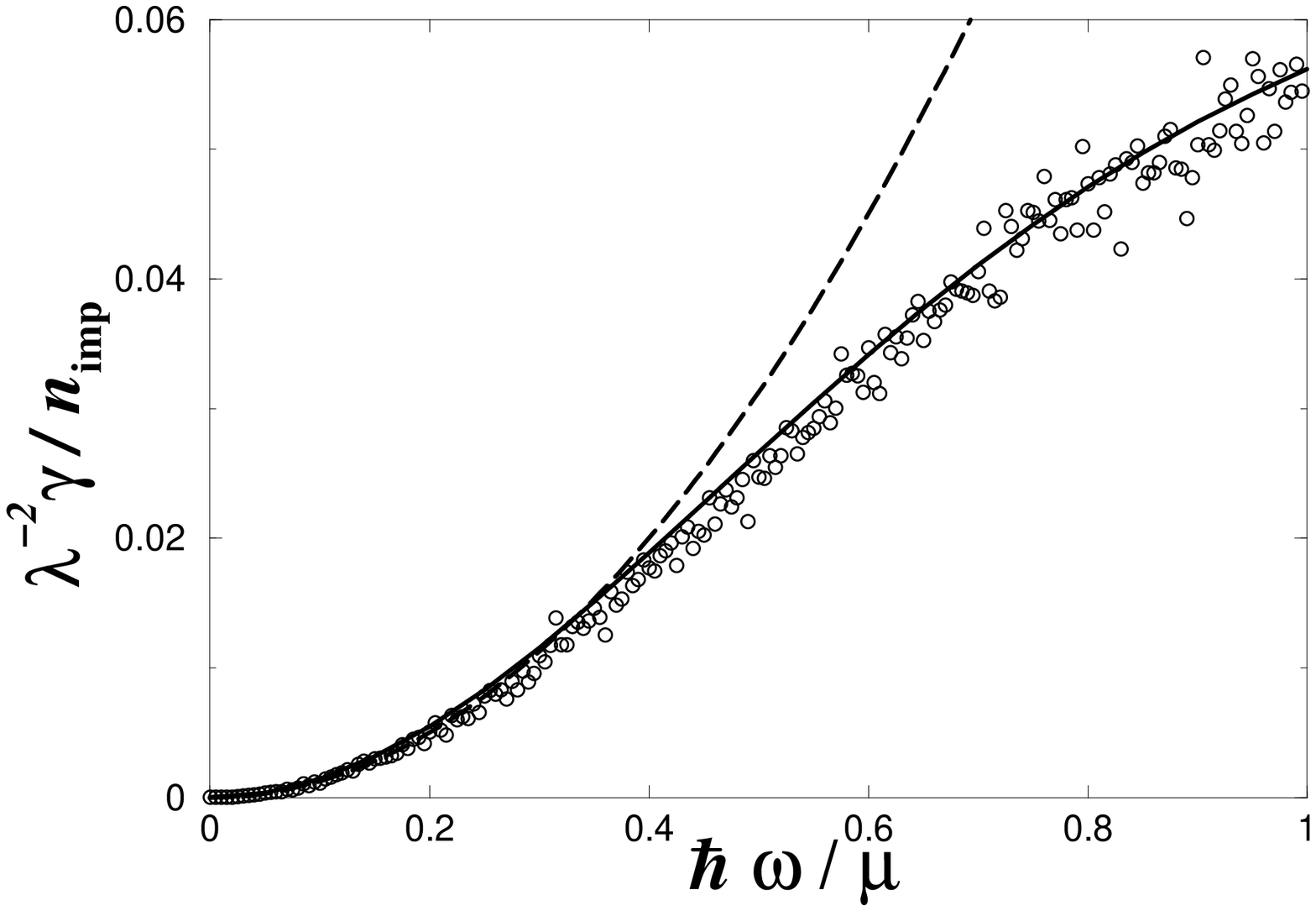}

Figure 5: {\sl $\gamma$ as a function of $\omega$ in rescaled units. The plot
	is drawn for $\lambda=0.2$ and $\xi\,n_{\rm imp}=0.03$. The dots are the
	results of the numerical simulation and the solid line is the analytical
	result from Eq.~(\ref{t7b}). The dashed line is the approximate result
	(\ref{t8}).}

\

 Formula (\ref{t8}) is interesting because it is identical to the first term
of expansion (\ref{h17}) which has been obtained in Section \ref{hydro} in a
completely different framework, and this permits to bridge the gap between the
hydrodynamical approach and the present transfer matrix method. As just
mentioned, formula (\ref{t8}) is restricted to small values of $\lambda$, but
the approach of Section \ref{hydro} is similarly limited to the domain $U_{\rm
typ}\ll \mu$. Also, Eq. (\ref{h17}) is restricted to small values of
$\omega\delta$. But in the present case $\omega\delta=\xi n_{\rm
imp}\,\lambda^2 (\hbar\omega/\mu)$ is very small, even if $\hbar \omega
\sim\mu$, so the restriction $\omega\delta\ll 1$ turns out to be of no
practical importance here. Also, the results obtained in the present Section
correspond to a potential with $\langle U\rangle=\lambda\,\mu\,\xi\, n_{\rm
imp}\ne 0$. However, the comparison with the results of Section \ref{hydro} is
possible with the rule given at the end of this Section for treating the case
of a potential with non zero mean. In this case the first term of expansion
(\ref{h17}) modifies to $\gamma=\frac{1}{8}\lambda^2 n_{\rm imp}
(\hbar\omega/\mu)^2(1-\lambda\xi n_{\rm imp})^{-3}$. The correcting term
$(1-\lambda\xi n_{\rm imp})^{-3}$, due to the non zero average of the
potential, gives an undetectable modification of the result (the relative
difference with (\ref{t8}) is of order 0.18 \% in the case of Figure 5).

\section{Discussion and conclusion}\label{conclusion}

  In this paper we have studied Anderson localization of elementary
excitations in a 1D BEC system. Emphasis has been put on the determination of
the localization length which has been determined in Section \ref{hydro} using
the ``phase formalism'' in the hydrodynamical approach (valid for
$\hbar\omega\ll\mu$) and in Section \ref{transfer}, using a transfer matrix
approach valid in the whole energy domain in the 1D mean field regime
(provided $n_{\rm imp}\xi\ll 1$). Results from the two approaches match within
the appropriate limit. The hydrodynamic approach has the advantage of being
able to deal with a large range of linear densities, ranging from the low
density Tonks-Girardeau regime to the high density transverse Thomas-Fermi
regime. In particular the puzzling absence of localization at low energy in
the Tonks-Girardeau limit deserves further studies.

Our findings can be tested in realistic experimental setups. Up to now, 3
experiments, lead at Firenze, Orsay and Hannover, have been done which all use
similar configurations \cite{Lye05,Cle05,Sch05}. Each of these experiments
involves an elongated cigar shaped condensate in a magnetic trap with an
optical speckle pattern creating the disordered potential \cite{Han}. The
experimental random potential has a non zero mean value, and the experiments
are done in the transverse Thomas-Fermi regime. We can thus study localization
in this configuration using (for excitations of energy small compared to the
chemical potential $\mu$) the above hydrodynamical approach of Section
\ref{hydro} adapted as explained at the end of this Section (replacing in all
the formulas $\mu$ by $\mu'_1=\frac{\mu}{2}-\langle U \rangle$,
$c_1'=(\mu_1'/m)^{1/2}$, etc...). One writes $U(x)=\langle U \rangle +
U_1(x)$. The auto-correlation $\langle U_1(x)U_1(0)\rangle$ has a typical
range $r_c$ which is in all the cases much larger than the healing length
$\xi$: $r_c=20\,\mu$m and $\xi=0.35 \;\mu$m in the Firenze experiment;
$r_c=5.2\;\mu$m and $\xi=0.16\;\mu$m in the Orsay experiment \cite{xi};
$r_c\approx 7\;\mu$m and $\xi=0.3\;\mu$m for $N=8\times10^4$ atoms at
Hannover. The condition (\ref{h3}) is fulfilled provided the pulsation
$\omega$ of the excitations is much lower than $2\pi c'_1/r_c$ (which, for
instance is equal to $2\pi\times 340$ Hz for $\langle U\rangle/\mu=0.2$ in the
Orsay experiment). In this regime, the potential can be approximated by a
white noise with a coefficient $D$ and a correlation radius $r_c$ defined by
\begin{equation}\label{c1}
\left(\frac{\hbar^2}{m}\right)^2 D = \int_\RR \langle U_1(x)U_1(0)\rangle \, 
{\rm d}x
=\langle U\rangle^2 \, r_c \; .
\end{equation}
Following the procedure explained in Section \ref{hydro} this leads to
\begin{equation}\label{c2}
L_{\rm loc}= \frac{\xi^2}{r_c} 
\left(\frac{\mu}{\langle U\rangle}\right)^2
\left(\frac{\mu}{\hbar\omega}\right)^2
\left(1-\frac{2\,\langle U\rangle}{\mu}\right)^3 \; .
\end{equation}
The experimental configuration which is closer to the one
considered in the present paper is the one of the Firenze group \cite{Lye05}
which has studied elementary excitations of an elongated condensate in
presence of a speckle pattern. The discrete excitation modes in elongated
systems are similar to the continuous ones of infinite systems 
we have described in the present article only in the
case of high quantum numbers (see, e.g., Ref. \cite{Pit03}, chap. 12).
Unfortunately, only the low lying dipole and quadrupole modes have been
studied in Ref. \cite{Lye05}. We nonetheless discuss this experiment using our
results, keeping in mind that we can only provide rough orders of magnitude.

The data of the Firenze group are presented in a way more easily analyzed
within the model of random delta peaks of Section \ref{transfer}. However, in
the regime where Eq.~(\ref{c2}) is valid, all models are expected to yield the
same result, as argued in Section \ref{hydro} and verified in Section
\ref{transfer}. The only relevant parameter being the parameter $D$, or
equivalently $r_c$ [which is related to $D$ by (\ref{c1})]. Within the model
of random $\delta$-peaks one has $\langle U\rangle=\lambda\, \mu\, \xi\,
n_{\rm imp}$ and $D=n_{\rm imp} (\lambda/\xi)^2$ yielding $r_c=n_{\rm
imp}^{-1}=20\;\mu$m \cite{Lye05}. The chemical potential in the Firenze
experiment is $\mu=1$ kHz and the excitations considered are the dipole
($\nu_1=8.74$ Hz) and quadrupole ($\nu_2=13.8$ Hz). For a disorder such that
$\langle U\rangle/\mu=0.1$ (which is typical in this experiment) the dipole
excitation corresponds to a localization length $L_{\rm loc}^1=4.1$ mm,
whereas for the quadrupole one gets $L_{\rm loc}^2=1.6$ mm ($L_{\rm
loc}^2=\frac{2}{5} L_{\rm loc}^1$ since $\nu_2/\nu_1=\sqrt{5/2}$). We also
note that higher excited modes having frequency
$\nu_n=\frac{\nu_1}{2}\sqrt{n(n+3)}$ \cite{Fli97,Str98} have lower
localization lengths: $L_{\rm loc}^n=\frac{4}{n(n+3)}\,L_{\rm loc}^1$. $L_{\rm
loc}^n$ becomes comparable with the the typical axial size of the condensate
(110 $\mu$m) for $n\sim 10$ \cite{speckle}.

A precise plot of the oscillations of a dipole mode is presented in Ref.
\cite{Lye05} in the case $\langle U\rangle/\mu=0.06$ which corresponds to a
limit we can address using Eq. (\ref{c2}) \cite{caveat}. An experimental
estimate of the value of the localization length can be obtained by fitting
the experimental data with a sinusoidal oscillation at frequency $\nu_1$ with
a dam\-ping $\exp\{-2\, X(t)/L_{\rm loc}^{\rm exp}\}$, where $X(t)=4 \Delta
\nu_1 t$ is the distance traveled by the dipole mode for an oscillation of
maximal amplitude $\Delta$. From the data presented in Ref. \cite{Lye05} we
obtain $L_{\rm loc}^{\rm exp}\simeq 1.7$ mm. This does not agree with the
value $L_{\rm loc}^1=15$ mm obtained from Eq. (\ref{c2}) in the case $\langle
U\rangle/\mu=0.06$, but we recall that we do not expect the dipole mode to be
equivalent to an excitation of an infinite system. Thus, the damping observed
in the Firenze experiment \cite{Lye05} cannot be accounted for by a model of
infinitely long condensate with no axial trapping. Quantitative theoretical
description of this experiment should take the axial trapping fully into
account. We nonetheless hope that the experimental study of higher excited
modes could directly confirm the result (\ref{c2}).

It is also interesting to discuss the expected localization length in the
Orsay experiment \cite{Cle05}, where the properties of the random potential
are well characterized. In this experiment, the potential is Poisson
distributed with a mean value $\langle U\rangle$ which is a fraction of the
chemical potential ($\mu=4.47$ kHz). Taking $\langle U\rangle/\mu=0.2$, and
for instance $\omega=\omega_z=2\pi\times 6.7$ Hz (corresponding to the dipole
excitation) one obtains $L_{\rm loc}=11.8$ mm. Besides, if one in able to
ge\-ne\-rate excitations with $\omega\simeq 6\times \omega_z$, one still
remains in the hydrodynamical regime and the above value of $L_{\rm loc}$ is
decreased by a factor 36, becoming of the order of the axial size of the
condensate (300 $\mu$m in the Orsay experiment \cite{Cle05}). We recall that
the present approach does not strictly apply for low lying excitations of a
trapped condensate, but it is nevertheless interesting to get an estimation of
the typical length scale for observing Anderson localization experimentally.

We note that the Firenze \cite{For05} , Orsay \cite{Cle05} and Hannover
\cite{Sch05} groups observed a saturation of the expansion of a condensate in
a disordered potential. In the 3 experiments this phenomenon has been
interpreted (see also \cite{Mod06,San06}) as being due to the trapping of the
wings of the condensate by the large peaks of the speckle potential, with no
relation to Anderson localization. We hope that in the near future, new
experiments will be able to directly address Anderson loca\-li\-za\-tion of
elementary excitations in transversely confined Bose-Einstein condensates, in
configurations corresponding to the scenario analyzed in the present work. In
this case, our study indicates that localization is more easily achieved for
excitations of energy of order $\mu$ (see Fig. 1) created for instance through
Bragg spectroscopy \cite{Ric03}. This range of energy is out of the
hydrodynamical regime presented in Section \ref{hydro}, but the approach of
Section \ref{transfer} allows to get a quantitative estimate of $L_{\rm loc}$
in this case (for the 1D mean field regime).

\

It is a pleasure to acknowledge fruitful discussions with L. Pastur, G.
Shlyapnikov and C. Texier. This work was supported by the Minist\`ere de la
Recherche (Grant ACI Nanoscience 201), by the ANR (grants
ANR--05--Nano--008--02 and ANR--NT05--2--42103) and by the IFRAF Institute.
Laboratoire de Physique Th\'eorique et Mod\`eles Statistiques is Unit\'e Mixte
de Recherche de l'U\-ni\-ver\-si\-t\'e Paris XI et du CNRS, UMR 8626.

\appendix
\section{Appendix: Density of state within the phase formalism}\label{dos}
In this Appendix we briefly demonstrate Eq.~(\ref{h19}) following a similar
demonstration in Ref. \cite{Lif88}. We first demonstrate  
that the phase $\theta$ defined in (\ref{h6}) is a monotonic function of
$\omega$. This can be shown by introducing the auxiliary variable
$y=-c_0 z/\omega$. Expressing (\ref{h10}) in terms of the variable $y$,
differentiating with respect to $\omega$ and then integrating the resulting
equation one gets
\begin{equation}\label{dos1}
\frac{\partial y}{\partial \omega}=
\frac{2\omega}{c_0^2}\int_0^x\!\!\!\!{\rm d}x'\,y^2(x')
\exp\{\frac{2\omega^2}{c_0^2}\int_{x'}^xy(x''){\rm d}x''\} \; > \; 0 \; .
\end{equation}
Thus $z=-\omega y/c_0$ is a decreasing function of $\omega$, and $\theta$ is
an increasing function of $\omega$ (since $\theta$ is a continuous function
and $\partial z/\partial\theta=-1-z^2$).
$\theta$ verifies the equation
\begin{equation}\label{dos2}
\frac{c_0}{\omega}\,\frac{{\rm d}\theta}{{\rm d}x} =
1+\sin^2\theta \, \frac{U(x)}{\mu} \; .
\end{equation}
The fact that $\theta$ is an increasing function of $\omega$ immediately 
implies that the number of eigenmodes [solutions of (\ref{h2}) or equivalently
of (\ref{h5b})] with pulsation between $0$ and $\omega$,
verifying the boundary condition $\cot\theta(0)=\theta_0$ and
$\cot\theta(L)=\theta_L$ 
coincides with the number of pulsations $\omega'\in[0,\omega]$ for which the
accumulated phase
$\theta(\omega',L)$ as determined by (\ref{dos2}) with the initial condition
$\theta(\omega',0)=\theta_0$ verifies
$\theta(\omega',L)=\theta_L+m\,\pi$. This number is equal to
\begin{equation}\label{dos3}
{\rm E}\left[\frac{\theta(\omega,L)-\theta(0,L)}{\pi}\right] =
{\rm E}\left[\frac{\theta(\omega,L)-\theta_0}{\pi}\right] \; ,
\end{equation}
\noindent where ${\rm E}(x)$ denotes the integer part of $x$. Passing to the
limit $L\to\infty$ and allowing for the fact the number of states is a self
averaging quantity one obtains
\begin{equation}\label{dos4}
N(\omega)=\lim_{L\to\infty} \frac{\langle\theta(\omega,L)\rangle}{\pi\,L} \; .
\end{equation}
But the number (\ref{dos3}) is also seen to be the number of times
where the variable $\theta$ equals zero modulo $\pi$ in the interval $[0,L]$.
This stems from the fact that $\theta(x)$ can change interval
$[n\pi,(n+1)\pi]$ only toward a higher interval and cannot go backward to a
lower interval, because, as seen from (\ref{dos2}), $\left.{\rm d}\theta/{\rm d}
x\right|_{\theta=n \pi}=\omega/c_0>0$. One may thus write
\begin{equation}\label{dos5}
N(\omega)=\lim_{L\to\infty}
\frac{\omega}{c_0\,L}\int_0^L\!\!{\rm d}x\;Q^{\rm red}(0;x) \; ,
\end{equation}
\noindent where 
\begin{equation}\label{dos6}
Q^{\rm red}(\theta;x)=\sum_{n\in\ZZ} 
\Big\langle\delta\big(\theta(x)-n\,\pi-\theta\big)\Big\rangle \; ,
\end{equation}
\noindent is the probability density for the reduced phase. Owing to the fact
that $Q^{\rm red}$ reaches a stationary (i.e., $x$ independent) 
distribution $Q^{\rm red}_{\rm st}(\theta)$, (\ref{dos5}) yields
\begin{equation}\label{dos7}
N(\omega)=\frac{\omega}{c_0}\, Q^{\rm red}_{\rm st}(0)= 
\frac{\omega}{c_0} \lim_{z\to\infty} (1+z^2)\,P_{\rm st}(z) \; ,
\end{equation}
\noindent where the last equality follows from the relation $z=\cot\theta$.
The explicit expression (\ref{h14}) of $P_{\rm st}$ evaluated at
large $z$ then yields the desired result (\ref{h19}).

\section{Appendix: Transmission through a single delta peak}\label{1impurity}
In this Appendix we determine the transmission and reflexion amplitude of an
elementary excitation of energy $\hbar\,\omega$ incident from the left on a
delta-like impurity located at $x=0$. These coefficients have already been
obtained by Kagan {\it et al.} in the case of a barrier of finite width
\cite{Kag03}. In the present case the impurity interacts with the atoms
forming the condensate via a potential $\lambda\,\mu\,\xi\,\delta(x)$ with
$\lambda>0$. The condensate is deformed near the impurity and the order
parameter reads
\begin{eqnarray}\label{b1}
\psi(x)& = & \tanh(|x/\xi|+a)\;\; , \quad {\rm with} \nonumber \\
a & = & \frac{1}{2} \,
\sinh^{-1}\left(\frac{2}{\lambda}\right) \; .
\end{eqnarray}
This form of $\psi(x)$ corresponds to two portions of black solitons matched
together at $x=0$ in order to satisfy the condition
$\xi\,[\psi'(0^+)-\psi'(0^-)]=2\,\lambda\,\psi(0)$. Far from the impurity (at
$x\to\pm\infty$), the background is not perturbed and an elementary excitation
of energy $\hbar\,\omega$ has a wave vector $q$ such that $\omega=c_0
q(1+q^2\xi^2/4)^{1/2}$, and is described by $(u(x),v(x))=\exp(i q
x)(u_\omega,v_\omega)$ where -- by Eq. (\ref{e4}) --
the constants $u_\omega$ and $v_\omega$ are
related by
\begin{equation}\label{b2}
\left(\frac{\xi^2q^2}{2}+1-\frac{\hbar\omega}{\mu}\right)u_\omega+v_\omega=0\; .
\end{equation}
The background is deformed near the impurity [as described by (\ref{b1})], and
in this region the form of the wave function of the elementary excitation is
affected in a non trivial manner. However, one still has an analytical
description of the excitations around the stationary profile
(\ref{b1}) because the expression of the
excitation around a soliton is known (it is given by the squared Jost
functions of the inverse pro\-blem \cite{Chen98}, see also Appendix A of Ref.
\cite{Bil05a}). Thus one can write the appropriate incoming, transmitted and
reflected modes of the problem. It is important however to realize that the
system has also evanescent modes localized around the impurity \cite{Kag03}.
More specifically, the scattering process of an excitation of energy
$\hbar\omega$ incident from $-\infty$ is described by
\begin{equation}\label{b3}
\Xi^{(-)}(x)=A_{\rm inc} \; \Xi^*_q(-x)+
A_{\rm ref} \; \Xi_q(-x)
+A^{(-)}_{\rm eva} \;\Xi_{i p}(-x)
 \; ,
\end{equation}
\noindent when $x<0$, and 
\begin{equation}\label{b4}
\Xi^{(+)}(x)=A_{\rm tra} \; \Xi_q(x)+A^{(+)}_{\rm eva}\; \Xi_{i p}(x) \; ,
\end{equation}
\noindent when $x>0$. The indexes ``inc'', ``ref'', ``tra'' and ``eva''
correspond respectively to incident, reflected, transmitted and evanescent
channels. The expression of $\Xi_k(x)$ ($k=q$ or $ip$) 
in (\ref{b3}) and (\ref{b4}) is
\begin{equation}\label{b6}
\Xi_k(x)=
e^{i k x} \left(
\begin{array}{c}
\left[\frac{k \xi}{2}+\frac{\omega}{c_0 k}+i\tanh(\frac{x}{\xi}+a)\right]^2 \\
\left[\frac{k \xi}{2}-\frac{\omega}{c_0 k}+i\tanh(\frac{x}{\xi}+a)\right]^2
\end{array}\right)\; ,
\end{equation}
and the quantities $q$ and $p$ are wave vectors related to $\omega$ by
\begin{eqnarray}\label{b5}
q\, \xi & = & \sqrt{2}
\left\{\sqrt{(\hbar\omega/\mu)^2+1}-1\right\}^{1/2} \; ,
\nonumber \\
p\, \xi & = & \sqrt{2}
\left\{\sqrt{(\hbar\omega/\mu)^2+1}+1\right\}^{1/2} \; .
\end{eqnarray}

The wave functions defined in Eqs. (\ref{b3}) and (\ref{b4}) are the most
general solutions of (\ref{e4}) corresponding to an elementary excitation of
energy $\hbar\omega$ incoming from the left and scattering on a potential
$U(x)=\lambda\,\xi\,\mu\,\delta(x)$. In particular, the incident, transmitted
and reflected components of (\ref{b3},\ref{b4}) all verify (\ref{b2}) far from
the impurity. The assumption $n_{\rm imp}\xi\ll 1$ made in Section
\ref{transfer} ensures that the evanescent mode $\Xi_{ip}$ does not reach the
nearest impurity \cite{evan}. This is the reason why the scattering on
potential (\ref{t1}) can be described via a transfer matrix approach using
only $2\times 2$ matrices.

The matching at $x=0$ corresponds to $\Xi^{(-)}(0)=\Xi^{(+)}(0)$ and
$\left.{\rm d}\Xi^{(+)}/{\rm d}x\right|_0 - \left.{\rm d}\Xi^{(-)}/{\rm
d}x\right|_0=2\,\lambda\,\xi^{-1} \Xi(0)$. This yields a system of 4 linear
equations determining the coefficients $A_{\rm ref}$, $A_{\rm tra}$,
$A^{(-)}_{\rm eva}$ and $A^{(+)}_{\rm eva}$ in terms of $A_{\rm inc}$. A
tedious but straightforward computation yields
\begin{equation}\label{b7}
t_\lambda=\frac{A_{\rm tra}}{A_{\rm inc}}=
\frac{1}{2}
\left[\frac{2+iq\xi\tanh(2a)}{-2+iq\xi\tanh(2a)}+
\frac{\Delta^*}{\Delta}\right] \; ,
\end{equation}
and
\begin{equation}\label{b8}
r_\lambda=\frac{A_{\rm ref}}{A_{\rm inc}}=
\frac{1}{2}
\left[\frac{2+iq\xi\tanh(2a)}{-2+iq\xi\tanh(2a)}-
\frac{\Delta^*}{\Delta}\right] \; ,
\end{equation}
\noindent where
\begin{eqnarray}\label{b9}
& \Delta & =  4\left(\frac{\hbar\omega}{\mu}+
2\,i\,\tanh^2a\right)\,\sqrt{\left(\frac{\hbar\omega}{\mu}\right)^2+1} 
 \\
      & + & 
2\,\xi(p+iq)\tanh a\left[\frac{2\hbar\omega}{\mu}+i(1+\tanh^2a)\right]\; .
\nonumber
\end{eqnarray}
The transmission probability $T_\lambda=|t_\lambda|^2$ has the asymptotic form
$T_\lambda \simeq 1 - \lambda^2
\mu/(2\hbar\omega)$ when $\omega\to\infty$, and in the opposite small energy
limit ($\hbar\,\omega\ll\mu$) one has
\begin{eqnarray}\label{b10}
	T_\lambda & \simeq& 1-\left(\frac{\hbar\,\omega}{2\,\mu}\right)^2
\left[1-\frac{2}{\tanh a}+\tanh(2a)\right]^2 \nonumber \\
& \APPROX{\lambda\to 0}& 
1 - \left(\frac{\lambda\,\hbar\,\omega}{2\,\mu}\right)^2 \; .
\end{eqnarray}
A typical behavior of $T_\lambda$ as a function of $\omega$ is plotted in Fig.
6. The transmission probability is 1 at small frequency. This anomalous
behavior of the transmission at small energy has already been noticed in Ref.
\cite{Kag03} in the case of a barrier of finite extend. It is also in
agreement with the findings of Ref. \cite{Bil05a} where a dark soliton with
velocity $v_{\rm sol}\to c_0$ (and thus reaching the limit where it becomes a
mere density perturbation, i.e., a phonon, which is an elementary excitation
with $q\to 0$) was shown to pass over an obstacle without radiating energy,
i.e., without reflection.

\

\includegraphics*[width=8cm]{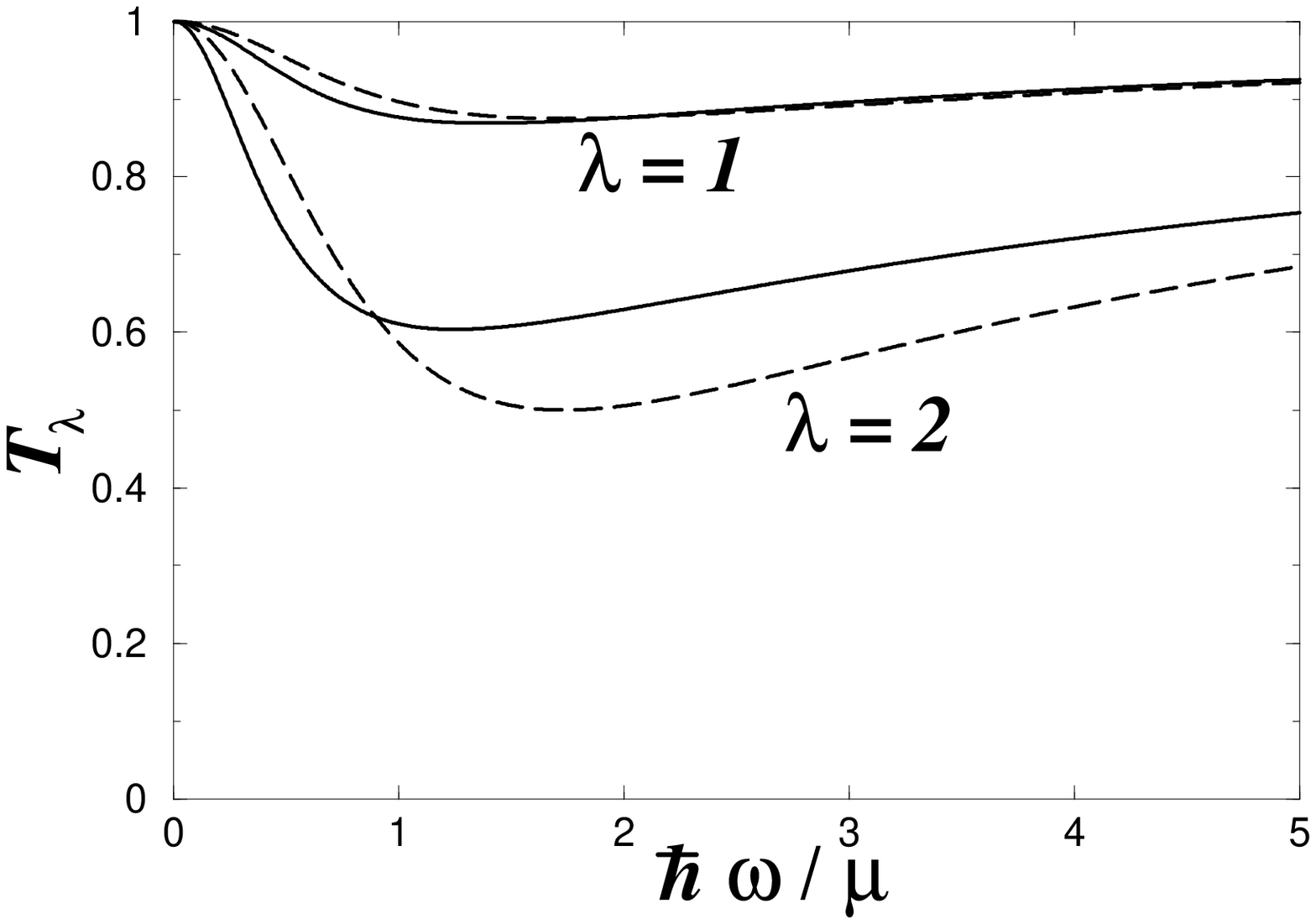}

Figure 6: {\sl Transmission probability $T_\lambda$ across a potential
	$U(x)=\lambda\,\mu\,\xi\,\delta(x)$ as a function of $\omega$ (in rescaled
	units). The solid line is the exact result (\ref{b7}) in the cases
	$\lambda=1$ and $\lambda=2$. The dashed line are the corresponding small
	$\lambda$ approximations (\ref{b11}).}

\

The exact formula for $T_\lambda$ [from (\ref{b7})] is compared on Figure 6 with
an approximation valid for all $\omega$ when $\lambda\ll 1$:
\begin{equation}\label{b11}
T_\lambda\simeq 1 - \frac{(\lambda\,\xi\,q/2)^2}{(\hbar\omega/\mu)^2+1} \; .
\end{equation}
It is seen on the Figure that this approximation is reasonably accurate already
when $\lambda= 1$. More precisely, for $\lambda=1$, the relative error
due to the use of Eq. (\ref{b11}) is lower than 3 \% (the error is maximum
around $\hbar\omega\simeq 0.7\mu$); for $\lambda=0.5$, this errors is lower
than 0.5 \%.


\begin{thebibliography}{99}
\bibitem{TFnote} Note that this is an {\it axial} Thomas-Fermi approximation
	which is done when the {\it transverse} wave-function is far form the
	Thomas-Fermi regime. When the transverse component gets also in the
	Thomas-Fermi regime, Eq. (\ref{h1}) still holds, but the form of the sound
	velocity in (\ref{h2}) is modified (as discussed at the end of Section 3).
\bibitem{rigor} A mathematically more rigorous presentation would be to have
	made no hypothesis on the type of randomness of $U(x)$ so far, and only at
	the present point to make the assumption --based on physical motivations--
	that $\langle\eta\rangle=0$ and $\langle\eta(x)\eta(0)\rangle=
	\xi^4D\,\delta(x)$ which is equivalent to (\ref{h4}) if $\eta(x)=U(x)/\mu$.
\bibitem{beware} This makes sense when the typical value $U_{\rm typ}$ of
$|U(x)|$ is much smaller than $\mu$, since in this case $c(x)/c_0\simeq 1$,
and from (\ref{h23}) $\alpha(x)$ is thus roughly similar to $\delta n(x)$.
\bibitem{avdist} This condition is valid even when $\lambda\to\infty$. In this
case, each impurity induces a deformation of the background which is a black
soliton centered at the position of the impurity, with typical extend $\xi$.
The nearest impurity is typically located at distance $n_{\rm imp}^{-1}$
and thus does not see the perturbation induced by its neighbor if
$n_{\rm imp}\xi\ll 1$.
\bibitem{fraction} The exact fraction of lengths inferior to $\xi$ is
	$\int_0^\xi P(\ell)d\ell= 1 - \exp\{-\xi n_{\rm imp}\}$, which is
	approximatively equal to $\xi\,n_{\rm imp}$ when $\xi\,n_{\rm imp} \ll 1$.
\bibitem{Han} The Hannover experiment has the possibility of an additional
	optical lattice.
\bibitem{xi} The conventions we use in the present work is slightly different
	from the ones of Ref. \cite{Cle05}~: we note $r_c$ what is denoted by
	$\Delta z$ in this reference, and there is a difference by a factor
	$\sqrt{2}$ in our definition of $\xi$ and in the one used in Ref.
	\cite{Cle05}.
\bibitem{speckle} A possible way to further decrease the localization length
would be to decrease the typical peak spacing of the speckle pattern.
\bibitem{caveat} The damping of the dipole oscillations are also studied for
other amplitude of disorder in Ref. \cite{Lye05}: $\langle U\rangle/\mu=0.4$
and 1, but we cannot address these regimes, Eq. (\ref{c2}) being only valid if 
$2\,\langle U\rangle/\mu\ll 1$.
\bibitem{evan} This is due to the exponential decrease
$\exp\{-p|x|\}$ of these modes, and to the fact that, from (\ref{b5}), the
wave vector $p$ is approximatively $2\,\xi^{-1}$ when $\omega\to 0$, and
$[2\,\omega/(\xi\,c_0)]^{1/2}$ when $\omega\to\infty$. Thus, $\exp\{-p/n_{\rm
imp}\}$ is always small.
\end{thebibliography}
\end{document}